\documentclass[nohyper]{JHEP3}
\usepackage{cite,epsfig,amssymb,amsmath}

\newcommand{\mycomm}[1]{\hfill\break $\phantom{a}$\kern-3.5em{\tt===$>$ \bf #1}\hfill\break}
\newcommand{\mycommA}[1]{\hfill\break $\phantom{a}$\kern-3.5em{\tt   $>$ \bf #1}\hfill\break}

\newcommand{\be}{\begin{equation}}
\newcommand{\ee}{\end{equation}}
\newcommand{\ba}{\begin{eqnarray}}
\newcommand{\ea}{\end{eqnarray}}

\catcode`\@=11 
\def\lsim{\mathrel{\mathpalette\@versim<}}
\def\gsim{\mathrel{\mathpalette\@versim>}}
\def\@versim#1#2{\vcenter{\offinterlineskip
        \ialign{$\m@th#1\hfil##\hfil$\crcr#2\crcr\sim\crcr } }}
\catcode`\@=12 

\title{On threshold resummation beyond leading $1-x$ order}

\author{G. Grunberg\\
        Centre de Physique Th\'eorique,  \'Ecole
Polytechnique, CNRS,\\
        91128 Palaiseau Cedex, France\\
        E-mail: \email{georges.grunberg@pascal.cpht.polytechnique.fr}}
        
  \author{V. Ravindran\\
  Harish-Chandra Research Institute,\\
  Chhatnag Road, Jhunsi,\\
  Allahabad 211 019, India \\
        E-mail: \email{ravindra@mri.ernet.in}}

\abstract{We check against exact finite order three-loop results  for the non-singlet $F_2$ and $F_3$  structure functions the validity  of a class of momentum space  ansaetze for threshold resummation at the next-to-leading order in $1-x$,  which generalize  results previously obtained in the large-$\beta_0$ limit. We find that the ansaetze do not work exactly, pointing towards an obstruction to threshold resummation at this order, but still yield correct results at the leading logarithmic  level for each color structures, as well as at the next-to-next-to-leading logarithmic  level for the specific $C_F^3$ color factor. A universality of the leading logarithm contributions to the physical evolution kernels of $F_2$ and $F_3$ at the next-to-leading order in $1-x$ is observed.}


\preprint{ }

\begin{document}

\section{Introduction}
Threshold resummation, which deals with the resummation to all orders of perturbation theory of the large logarithmic
corrections  arising from the incomplete cancellation of  soft and collinear gluons at the edge of phase
space, is by now a well developed topic \cite{Sterman:1986aj,Catani:1989ne} in perturbative QCD. Recently, some renewed  interest has been expressed \cite{Kramer:1996iq,Grunberg:2007nc,Kidonakis:2007ww,Laenen:2008ux,Laenen:2008gt} in the resummation of those logarithmically enhanced terms which are  suppressed by some power of $(1-x)$ for $x\rightarrow 1$ in momentum space (or by some power of $1/N$, $N\rightarrow\infty$ in moment space) with respect to the leading terms. In particular, in \cite{Grunberg:2007nc} a very simple form was obtained  
for the  structure of threshold resummation at all orders in $(1-x)$ in the large--$\beta_0$ limit in {\em momentum space}, and a straightforward generalization of the large--$\beta_0$ result to finite $\beta_0$ was suggested. The result in \cite{Grunberg:2007nc} was obtained by working at the level of the  momentum space physical evolution kernels (or `physical anomalous dimensions', see e.g. \cite{Furmanski:1981cw,Grunberg:1982fw,Catani:1996sc,Blumlein:2000wh,vanNeerven:2001pe}), which are infrared and collinear safe quantities describing the physical scaling violation, where the structure of the result appears to be particularly transparent. The purpose of this note is to check the finite $\beta_0$  conjecture of \cite{Grunberg:2007nc} by comparing with the three-loop calculations of \cite{Vermaseren:2005qc,Moch:2008fj} for the $F_2$ and $F_3$ non-singlet structure functions. It is found  that the conjecture in the simplest form (section 2) suggested in \cite{Grunberg:2007nc} does not actually work, neither do two other plausible generalizations (sections 3 and 4). We conclude, in agreement with the analysis in \cite{Laenen:2008gt}, that threshold resummation probably does not work exactly at the next-to-leading order in $1-x$. Neverheless, we show these ansaetze do suggest some correct predictions, both at the leading logarithmic (LL) order (for each color factor separately) at two and three loop, and at next-to-next-to-leading logarithmic (NNLL) order for the peculiar color factor $C_F^3$ at three loop. A comparaison with the closely related work of \cite{Laenen:2008ux} is also performed in section 5, and in the conclusion (section 6) we comment on the possible structure of a threshold resummation violating piece. The results
of some calculations involving convolutions are presented in more details in four appendices.

\section{The conjecture}

The scale--dependence of the (flavour non-singlet) deep inelastic ``coefficient function''
${\cal C}_2(x, Q^2,\mu^2_F)$ corresponding to the non-singlet $F_2(x,Q^2)$ structure function ($F_2(x,Q^2)/x={\cal C}_2(x, Q^2,\mu^2_F)\otimes q_{2}(x,\mu^2_F)$, where $ q_{2}(x,\mu^2_F)$ is the quark distribution)
 can be expressed in terms of ${\cal C}_2(x, Q^2,\mu^2_F)$ itself, yielding the following evolution equation
(see e.g.~Refs.~\cite{Grunberg:1982fw,Catani:1996sc,vanNeerven:2001pe}):
\begin{equation}
\label{F_2_evolution}
\frac{d{\cal C}_2(x,Q^2,\mu^2_F)}{d\ln Q^2}\,=\,\int_x^1 \frac{dz}{z}\,K(x/z,Q^2)\,{\cal C}_2(z,Q^2,\mu^2_F)\ ,
\end{equation}
where $\mu_F$ is the factorization scale (we assume for definitness the $\overline{MS}$ factorization scheme is used).
$K(x,Q^2)$ is the momentum space \emph{physical evolution kernel}, or {\em physical anomalous dimension}; it is independent of the factorization scale and renormalization--scheme invariant. In \cite{Gardi:2007ma}, using standard results \cite{Sterman:1986aj,Catani:1989ne} of Sudakov resummation in moment space, the result for the leading contribution to this quantity in the $x\rightarrow 1$ limit was derived:
\begin{equation}
\label{K_x}
K(x,Q^2)\sim \frac{{\cal J}\left((1-x)Q^2\right)}{1-x}\,+\,\frac{d\ln \left({\cal F}(Q^2)\right)^2}{d\ln Q^2}\,\delta(1-x)\ , \end{equation}
where ${\cal J}(k^2)$, the `physical Sudakov anomalous dimension', is defined in eq.(\ref{standard-J-coupling}) below.
Eq.(\ref{K_x}) shows that threshold resummation takes a very simple form directly in {\em  momentum-space} when dealing with the physical evolution kernel: ${{\cal J}\left((1-x)Q^2\right)}/({1-x})$ is the leading term in the expansion of the physical momentum space  kernel $K(x,Q^2)$ in the \hbox{$x\to 1$} limit with  $(1-x)Q^2$ fixed, and all threshold logarithms are absorbed into the {\em single} scale  $(1-x)Q^2$. The  term  proportional to $\delta(1-x)$ is comprised of purely virtual corrections associated with the quark form factor ${\cal F}(Q^2)$. This term is infrared divergent, but  the singularity cancels exactly upon integrating over~$x$ with the divergence of the integral of ${{\cal J} \left((1-x)Q^2\right)}/({1-x})$ near $x\to 1$. 

\noindent To derive eq.(\ref{K_x}), one starts from the standard threshold resummation formula for the moment space  coefficient function

\begin{equation}{\hat {\cal C}}_2(Q^2,N,\mu^2_F)=\int_0^1dx\, x^{N-1}{\cal C}_2(x, Q^2,\mu^2_F)\ , \end{equation} 
namely, for $N\rightarrow\infty$: 

\begin{equation}
{\hat {\cal C}}_2(Q^2,N,\mu^2_F)\sim g(Q^2,\mu^2_F)\
\exp[E(Q^2,N,\mu^2_F)]\label{resum}\ ,
\end{equation} 
with the Sudakov exponent given by
\begin{equation}
E(Q^2,N,\mu^2_F)=\int_0^1 dx
\frac{x^{N-1}-1}{1-x}\left[\int_{\mu^2_F}^{(1-x)Q^2}\frac{dk^2}{k^2}
A\left(a_s(k^2)\right)+B\left(a_s((1-x)Q^2)\right)\right]\ ,\label{exponent}
\end{equation}
where ($a_s\equiv\frac{\alpha_s}{4\pi}$) 
\begin{equation}\label{cusp}
A(a_s)=\sum_{i=1}^\infty
A_ia_s^{i}
\end{equation} 
is \cite{Korchemsky:1993uz} the universal ``cusp'' anomalous dimension,
and
\begin{equation}\label{B-stan}
B(a_s)=
\sum_{i=1}^\infty B_i a_s^{i}
\end{equation}
is the standard final state ``jet function''  anomalous dimension, whereas
$g(Q^2,\mu^2_F)$
collects the residual constant (i.e. $N$-independent) terms not included in $E(Q^2,N,\mu^2_F)$. It should be noted that both $A(a_s)$ and $B(a_s)$ are
renormalization scheme-dependent quantities.
\noindent Taking the $\ln Q^2$ derivative of
eq.(\ref{resum}) we get the large-$N$ resummation formula
\cite{Forte:2002ni,Gardi:2002xm}
 for the moment space ``physical evolution kernel'' 
  ${\hat K}(Q^2,N)\equiv{d\ln{\hat {\cal C}}_2(Q^2,N,\mu^2_F)\over d\ln Q^2}$:

 \begin{equation}\label{scaling-violation}
{\hat K}(Q^2,N)\sim\int_{0}^1 dx{x^{N-1}-1
 \over
1-x} {\cal J}[(1-x)Q^2]+ H(Q^2)\,
\end{equation}
 where
\begin{equation} 
H(Q^2)={d\ln g(Q^2,\mu^2_F)\over d\ln
Q^2}\label{H-stan}\ ,
\end{equation} 
and

\begin{eqnarray} \label{standard-J-coupling}
{\cal J}(k^2)&= &A\left(a_s(k^2)\right)+{dB\left(a_s(k^2)\right)\over d\ln k^2}\\
&=&A\left(a_s(k^2)\right)+\beta\left(a_s(k^2)\right)\frac{dB\left(a_s(k^2)\right)}{da_s}\nonumber\ ,\end{eqnarray}
is a ``physical'' (i.e. scheme-independent) Sudakov anomalous dimension, depending upon the ``jet scale'' $(1-x)Q^2$ in eq.(\ref{scaling-violation}).
Merging together the $N$-independent $(-1)$ contribution sitting inside the integral in (\ref{scaling-violation}) with $H(Q^2)$, one arrives at:

\begin{equation}\label{scaling-violation-bis}
{\hat K}(Q^2,N)\sim\int_{0}^1 dx\, x^{N-1}\frac
  {{\cal J}[(1-x)Q^2]}{1-x}+\left[ H(Q^2)-\int_{0}^{Q^2}{dk^2\over k^2}
{\cal J}(k^2)\right]\,
\end{equation}
Inverting the moments in eq.(\ref{scaling-violation-bis}), and using the relation\cite{Friot:2007fd}
 \begin{equation}\label{conjecture-0}
 H(Q^2)-\int_{0}^{Q^2}{dk^2\over k^2}
{\cal J}(k^2)={d\ln \left({\cal
F}(Q^2)\right)^2 \over d\ln Q^2}\ ,
\end{equation}
one  finally obtains the corresponding momentum space relation eq.(\ref {K_x}).

\noindent Eq.(\ref{K_x}) gives a strong incentive to look for a systematic expansion for $x\rightarrow 1$ of $K(x, Q^2)$ in powers of $1-x$, or, more conveniently, in powers of
\begin{equation}
\label{r-DIS}
r\equiv\frac{1-x}{x}
\end{equation}
 at {\em fixed} jet mass 
\begin{equation}
\label{W}
W^2\equiv r\,Q^2\ .
\end{equation}
The simplest guess would be:
\begin{equation}
\label{r-expansion-DIS}
K(x,Q^2)=\frac{1}{r}\,{\cal J}\left(W^2\right)\,+\,\frac{d\ln \left({\cal F}(Q^2)\right)^2}{d\ln Q^2}\,\delta(1-x)+{\cal J}_{0}\left(W^2\right)\,+{\cal O}\left(r\right),
\end{equation}
where (barring the virtual contribution) the coefficients ${\cal J}\left(W^2\right)$ and ${\cal J}_{0}\left(W^2\right)$  are renormalization group and scheme invariant `effective charges' \cite{Grunberg:1982fw}, the physical `jet' Sudakov anomalous dimensions, functions of a {\em single} variable--the jet mass $W^2$, that can be computed order by order in $a_s(W^2)$.  This ansatz has been checked \cite{Grunberg:2007nc} in the large--$\beta_0$ limit\footnote{$K(x,Q^2)$  defined here  is $1/x$ $\times$ the $K(x,Q^2)$ as defined  in  \cite{Grunberg:2007nc}.}.
A more general ansatz \cite{Grunberg:2007nc} could be:
\begin{equation}
\label{r-expansion1-DIS}
K(x,Q^2)=\frac{1}{r}\,{\cal J}\left(W^2\right)\,+\,\frac{d\ln \left({\cal F}(Q^2)\right)^2}{d\ln Q^2}\,\delta(1-x)+\left[\bar{{\cal J}}_{0}\left(W^2\right)\,\ln(1-x)+{\cal J}_{0}\left(W^2\right)\right]\,+{\cal O}\left(r\ln^2 r\right).
\end{equation}
Eq.(\ref{r-expansion1-DIS}) involves an `explicit' $\ln(1-x)$ factor at ${\cal O}(r^0)$, as suggested by the expansion of the standard splitting function (see eq.(\ref{P}) below).
  It turns out that the ansatz eq.(\ref{r-expansion-DIS}), and even eq.(\ref{r-expansion1-DIS}), do not work at finite $\beta_0$. An alternative ansatz which involves {\em two} different scales beyond $1/r$ order is suggested below (eq.(\ref{r-expansion2-DIS})), but does not work either.

\section{Checking the ansatz}

1)\underline{ ${\cal O}(a_s^2)$ exact result}:
let us first give the exact result for $K(x, Q^2)$ as $x\rightarrow 1$ at ${\cal O}(a_s^2)$. One starts from the general relation  \cite{vanNeerven:2001pe}

\begin{equation}\label{K-loops}K(x,Q^2)=P(x,a_s)+\beta(a_s)(d_1(x)+d_2(x)\, a_s+d_3(x)\, a_s^2+...)\, \end{equation}
where $a_s\equiv a_s(Q^2)$,

 \begin{equation}P(x, a_s)=\sum_{i=0}^\infty P_i(x) a_s^{i+1}\label{split}\end{equation}
 is the standard splitting function,
\begin{equation}\beta(a_s)=\frac{d a_s}{d\ln Q^2}=-\beta_0\, a_s^2-\beta_1\, a_s^3-\beta_2\, a_s^4+...\label{beta}\end{equation}
is the beta function (with $\beta_0=\frac{11}{3}C_A-\frac{2}{3}n_f$), and $d_i(x)$'s are the expansion coefficients of the formal logarithmic derivative (in the sense of convolutions) $d\ln{\cal C}_2/da_s$. Namely, setting

\begin{equation}{\cal C}_2(x, Q^2,\mu^2_F=Q^2)=\delta(1-x)+\sum_{i=1}^\infty c_i(x)\, a_s^i\label{ci}\ ,\end{equation}
we have

\begin{eqnarray}d_1(x)&=&c_1(x)\label{di}\\
d_2(x)&=&2c_2(x)-c_1^{\otimes 2}(x)\nonumber\\
d_3(x)&=&3c_3(x)-3c_2(x)\otimes c_1(x)+c_1^{\otimes 3}(x)\nonumber\ , \end{eqnarray}
with $c_1^{\otimes 2}(x)=c_1(x)\otimes c_1(x)$, etc...
Expanding eq.(\ref{K-loops}) to ${\cal O}(a_s^2)$, one gets:

\begin{equation}
\label{K-2loop}
K(x,Q^2)=a_s\, P_0(x)+a_s^2[P_1(x)-\beta_0\, c_1(x)]+...\ .
\end{equation}
Now for $x\rightarrow 1$, we have:

\begin{equation}
\label{P0}
P_0(x)\sim \frac{A_1}{r}+B_1^{\delta}\, \delta(1-x)+C_1\ln(1-x)+D_1+...
\end{equation}
with  $A_1=4C_F$, $B_1^{\delta}=3C_F$,  $C_1=0$ and $D_1=0$,

\begin{equation}
\label{P1}
P_1(x)\sim \frac{A_2}{r}+B_2^{\delta}\, \delta(1-x)+C_2\ln(1-x)+D_2+...
\end{equation}
with \cite{Kodaira:1981nh} $A_2=(\frac{16}{3}-8\zeta_2)C_F\, C_A+\frac{20}{3}C_F\, \beta_0$ (where we have expressed for convenience $n_f$ in term of $\beta_0$ and $C_A$),
 $C_2=A_1^2$,  and  \cite{Dokshitzer:2005bf,Basso:2006nk} 
 
 \begin{equation}D_2=A_1(B_1^{\delta}-\beta_0)\label{D2}\ .\end{equation} 
 Moreover:
 
 \begin{equation}
\label{c1-usual}
c_1(x)\sim C_F[\frac{4\ln(1-x)-3}{1-x}-(9+4\zeta_2)\delta(1-x)-4\ln(1-x)+14+...]\ ,
\end{equation}
where the $\frac{\ln^p(1-x)}{1-x}$ terms should be interpreted from now on as $+$-distributions, which makes the coefficient of the $\delta(1-x)$ term finite.
Hence, in an  expansion in $1/r=1/(1-x)-1$,  we get (skipping the $\delta(1-x)$ term):

\begin{equation}
\label{c1}
c_1(x)\sim C_F[\frac{4\ln(1-x)-3}{r}+11+...]\ .
\end{equation}
We note that in an expansion in $1/r$, there is no logarithmic term at ${\cal O}(r^0)$ order in $c_1(x)$, a consequence of the fact \cite{Kramer:1996iq}  that the coefficients of the $\ln(1-x)/(1-x)$ and $\ln(1-x)$ leading logarithms in eq.(\ref{c1-usual}) are equal and opposite.
We thus get (skipping the $\delta(1-x)$ term)
\begin{eqnarray}
K(x,Q^2)&\sim&\frac{1}{r}[A_1\, a_s+a_s^2(-4C_F\beta_0\ln(1-x)+A_2+3C_F\beta_0)+...]\nonumber\\
&+&a_s^2[C_2\ln(1-x)+D_2-11C_F\beta_0]+...\label{r-expansion-2loop-DIS}
\end{eqnarray}
i.e.

\begin{eqnarray}
K(x,Q^2)&\sim&\frac{1}{r}[A_1\, a_s+a_s^2(-A_1\beta_0\ln(1-x)+A_2+3C_F\beta_0)+...]\nonumber\\
&+&a_s^2[A_1^2\ln(1-x)+A_1 B_1^{\delta}-(A_1+11C_F)\beta_0]+...\label{r-expansion-2loop-DIS-bis}
\end{eqnarray}
where we have replaced $4C_F$ by $A_1$ in the coefficient of the $\ln(1-x)$ term on the first line.

2) \underline{Ansatz}:

Let us first check the simpler ansatz eq.(\ref{r-expansion-DIS}). Using the well-known relations (following from renormalization group invariance):

\begin{equation}{\cal J}\left(W^2\right)=j_1a_s+a_s^2(-j_1\beta_0\ln(\frac{W^2}{Q^2})+j_2)+a_s^3[j_1\beta_0^2\ln^2(\frac{W^2}{Q^2})-(j_1\beta_1+2\beta_0\, j_2)\ln(\frac{W^2}{Q^2})+j_3]+...\label{j2}\end{equation}
and

\begin{equation}
{\cal J}_{0}\left(W^2\right)=j_{02}a_s^2+a_s^3(-2j_{02}\beta_0\ln(\frac{W^2}{Q^2})+j_{03})+....\label{j0-2loop}\end{equation}
 as well as the expansion:
 
 \begin{equation}\ln(\frac{W^2}{Q^2})=\ln r=\ln(1-x)+(1-x)+...\label{scale-expand}\ , \end{equation} 
 the ansatz  eq.(\ref{r-expansion-DIS})  yields at ${\cal O}(a_s^2)$ order:

\begin{eqnarray}
K(x,Q^2)\vert{ansatz}&\sim&\frac{1}{r}[j_1\, a_s+a_s^2(-\beta_0 j_1\ln(1-x)+j_2)+...]\nonumber\\
&+&a_s^2(j_{02}-j_1\beta_0)+...\label{r-expansion-2loop-DIS-ansatz1}
\end{eqnarray}
We note that  ${\cal J}_{0}\left(W^2\right)$ must be a ${\cal O}(a_s^2)$
quantity to match eq.(\ref{r-expansion-2loop-DIS}), where the subleading ${\cal O}(r^0)$ `next-to-eikonal' 
 term starts at 
${\cal O}(a_s^2)$ (this follows from the fact that $C_1=D_1=0$).  Comparing with eq.(\ref{r-expansion-2loop-DIS-bis}), one finds (as expected) that the leading ${\cal O}(1/r)$ term matches the corresponding one in (\ref{r-expansion-2loop-DIS-bis}), with the identifications:

\begin{eqnarray}\label{j1-j2}j_1&=&A_1\\
j_2&=&A_2+3\beta_0 C_F\nonumber\ ,\end{eqnarray} 
which implies (see eq.(\ref{standard-J-coupling})) $B_1=-3 C_F$.
Furthermore at the next-to-leading ${\cal O}(r^0)$ order, the non-logarithmic contribution in eq.(\ref{r-expansion-2loop-DIS-ansatz1}) matches the corresponding one in eq.(\ref{r-expansion-2loop-DIS-bis}) provided:

\begin{equation}j_{02}=A_1B_1^{\delta}-11C_F\beta_0\, .\label{j02-bis}\end{equation}
We also note the  $j_1\beta_0$ term on the second line of eq.(\ref{r-expansion-2loop-DIS-ansatz1}), which arises as a `remnant'  from the expansion of the  $\ln(W^2/Q^2)$ term occuring at ${\cal O}(a_s^2)$ (eq.(\ref{j2})) in the leading $1/r$ part of the ansatz, matches the $A_1\beta_0$ term in $D_2$ in eq.(\ref{r-expansion-2loop-DIS}).
Eq.(\ref{j02-bis}) shows that ${\cal J}_{0}\left(W^2\right)$ is not a total derivative (contrary to the situation which prevails  \cite{Grunberg:2007nc} at large $\beta_0$), but suggests that it can be written as  the sum of two components: 

\begin{equation}{\cal J}_{0}\left(W^2\right)={\tilde{\cal J}}_{0}\left(W^2\right)+\frac{d{\cal B}_0}{d\ln W^2}\label{J0-B}\end{equation}
where

\begin{equation}{\tilde{\cal J}}_{0}\left(W^2\right)={\tilde j}_{02} a_s^2+...\label{J0tilde}\ ,\end{equation}
with ${\tilde j}_{02}=A_1B_1^{\delta}$,
is non-leading at large  $\beta_0$, and

\begin{equation}{\cal B}_0(W^2)=b_1 a_s+...\label{B0}\end{equation}
with $b_1=11C_F$ (in agreement with the large $\beta_0$ analysis \cite{Grunberg:2007nc}). We note that $b_1$ corresponds to the contribution of the constant term in the  coefficient function (see eq.(\ref{c1})), whereas  ${\tilde j}_{02}$ is contributed by the splitting function.

However, there remains one obvious mismatch: the logarithmic contribution on the second line of eq.(\ref{r-expansion-2loop-DIS-bis}) is not accounted for by the ansatz eq.(\ref{r-expansion-DIS}).
This  mismatch suggests to try instead the ansatz eq.(\ref{r-expansion1-DIS}), with an explicit $\ln (1-x)$ term. Setting

\begin{equation}\bar{{\cal J}}_{0}\left(W^2\right)=\bar{j}_{02}a_s^2+a_s^3(-2\bar{j}_{02}\beta_0\ln(\frac{W^2}{Q^2})+\bar{j}_{03})+...\label{jbar0-2loop}\end{equation}
the ansatz  eq.(\ref{r-expansion1-DIS})  yields at ${\cal O}(a_s^2)$ order:

\begin{eqnarray}
K(x,Q^2)\vert{ansatz}&\sim&\frac{1}{r}[j_1\, a_s+a_s^2(-\beta_0 j_1\ln(1-x)+j_2)+...]\nonumber\\
&+&a_s^2[\bar{j}_{02}\ln(1-x)+(j_{02}-j_1\beta_0)]+...\label{r-expansion-2loop-DIS-ansatz2}
\end{eqnarray}
Indeed eq.(\ref{r-expansion-2loop-DIS-ansatz2}) matches (\ref{r-expansion-2loop-DIS-bis}) with the identification:

\begin{equation}\bar{j}_{02}=C_2=A_1^2\label{bar-j02}\end{equation}
together with eq.(\ref{j02-bis}). At ${\cal O}(a_s^2)$ the ansatz thus makes no prediction, but we note  that eq.(\ref{bar-j02}) is a consequence of the previously mentioned fact that  the coefficients of the $\ln(1-x)/(1-x)$ and $\ln(1-x)$ leading logarithms in $c_1(x)$
(eq.(\ref{c1-usual})) are equal up to a sign. Thus, assuming the parameter $j_1$  of the  ${\cal O}(1/r)$ part of the ansatz has been fixed as in eq.(\ref{j1-j2}) to correctly reproduce the $\ln(1-x)/(1-x)$ term in $c_1(x)$,  eq.(\ref{bar-j02}) guarantees the correct coefficient of the $\ln(1-x)$ term in $c_1(x)$ is obtained.

\noindent The determined value of $\bar{j}_{02}$ moreover reveals  an interesting pattern. It shows that\footnote{However, since $j_{02}\neq D_2$ (see eq.(\ref{D2}) and (\ref{j02-bis})),  the analoguous relation ${\cal J}_{0}\left(W^2\right)=D\left(a_s(W^2)\right)+{\cal O}(a_s^3)$ (where $D(a_s)=\sum_{i=2}^\infty D_i a_s^{i}$) is not realized. The reason is that, with the definition (\ref{W}) of the $W$ scale, the `remnant' $j_1\beta_0=4C_F\beta_0$ on the second line of eq.(\ref{r-expansion-2loop-DIS-ansatz1}), which arises from the factor of $x$ in the denominator of (\ref{r-DIS}), does not match the $b_1\beta_0=11C_F\beta_0$ term on the second line of eq.(\ref{r-expansion-2loop-DIS}).}:

\begin{equation}\bar{{\cal J}}_{0}\left(W^2\right)=C\left(a_s(W^2)\right)+{\cal O}(a_s^3)\ ,\label{analog1}\end{equation}
where 

\begin{equation}C(a_s)=\sum_{i=2}^\infty C_i a_s^{i}\ ,\label{C}
\end{equation}
is the coefficient of the ${\cal O}(\ln(1-x))$ term in the expansion  of the standard splitting function $P(x, a_s)=\sum_{i=0}^\infty P_i(x) a_s^{i+1}$ around $x=1$, namely \cite{Korchemsky:1988si} (see  eq.(\ref{P0}), (\ref{P1}) and (\ref{P2})):

\begin{equation}
\label{P}
P(x,a_s)=\frac{1}{r} A(a_s)+B^{\delta}(a_s)\, \delta(1-x)+[C(a_s)\ln(1-x)+D(a_s)]+{\cal O}(r \ln^2 r)\, .\end{equation}
Eq.(\ref{r-expansion1-DIS}) is clearly the analogue of eq.(\ref{P}). Moreover
eq.(\ref{analog1}) is quite analogous to eq.(\ref{standard-J-coupling}), relating the coefficients of the ${\cal O}(1/r)$ terms in $P(x, a_s)$ and $K(x, Q^2)$, which shows that 

\begin{equation}{\cal J}\left(W^2\right)=A\left(a_s(W^2)\right)+{\cal O}(a_s^2)\, .\end{equation}
Actually, since the physical kernel $K(x,Q^2)$ differs from the standard splitting function by a term proportional to the beta function (see eq.(\ref{K-loops})), one can even state that:

\begin{equation}\bar{{\cal J}}_{0}\left(W^2\right)=C\left(a_s(W^2)\right)+{d{\bar B}_0\left(a_s(W^2)\right)\over d\ln
W^2}\ ,\label{analog2}\end{equation}
with ${\bar B}_0(a_s)={\cal O}(a_s^2)$. We shall see in section 5 that this observation makes contact with the ansatz proposed in \cite{Laenen:2008ux}.

 On the other hand, a prediction does  arise at ${\cal O}(a_s^3)$ order. One finds:

\begin{eqnarray}
K(x,Q^2)\vert{ansatz}&\sim&\frac{1}{r}[j_1\, a_s+a_s^2(-\beta_0 j_1\ln(1-x)+j_2)\nonumber\\
& &+a_s^3(j_1\beta_0^2\ln^2(1-x)
-(j_1\beta_1+2\beta_0\, j_2)\ln(1-x)+j_3)+...]\nonumber\\
&+&a_s^2[\bar{j}_{02}\ln(1-x)+(j_{02}-j_1\beta_0)]+...\nonumber\\
&+&a_s^3[-2\bar{j}_{02}\beta_0\ln^2(1-x)
+(2\beta_0\,( j_1\beta_0-j_{02})+\bar{j}_{03})\ln(1-x)\nonumber\\
& &+(-j_1\beta_1-2\beta_0\, j_2+j_{03})]+...
\label{r-expansion-3loop-DIS-ansatz}
\end{eqnarray}
where the ${\cal O}(r^0)$ leading logarithms in the third and fourth lines of eq.(\ref{r-expansion-3loop-DIS-ansatz}) come from the expansion of the `explicit' $\ln (1-x)$ term in eq.(\ref{r-expansion1-DIS}):

\begin{equation}\bar{{\cal J}}_{0}\left(W^2\right)\,\ln (1-x)=a_s^2\,\bar{j}_{02}\ln(1-x)+ a_s^3[-2\bar{j}_{02}\beta_0\ln^2(1-x) +\bar{j}_{03}\ln(1-x)]+...\label{explicitlogs}\end{equation}
The ansatz eq.(\ref{r-expansion1-DIS}) thus predicts 

i) that the leading ${\cal O}(r^0)$ logarithm at ${\cal O}(a_s^3)$ in $K(x, Q^2)$ should be a double logarithm, and 

ii) that its coefficient should be $-2\bar{j}_{02}\beta_0=-2C_2\beta_0=-2A_1^2\beta_0$, which can be compared to the ${\cal O}(a_s^3)$ exact result.

3) \underline{${\cal O}(a_s^3)$ exact result}:
expanding eq.(\ref{K-loops}) to ${\cal O}(a_s^3)$ one gets:

\begin{eqnarray}
K(x,Q^2)&=&a_s\, P_0(x)+a_s^2[P_1(x)-\beta_0\, c_1(x)]\nonumber\\
&+&a_s^3[P_2(x)-\beta_1\, c_1(x)-\beta_0\, d_2(x)]+...\, .
\label{K-3loop}\end{eqnarray}
Using the $x\rightarrow 1$ expansion of the three loop splitting function:

 \begin{equation}
\label{P2}
P_2(x)\sim \frac{A_3}{r}+B_3^{\delta}\, \delta(1-x)+C_3\ln(1-x)+D_3+...
\end{equation}
where \cite{Moch:2004pa,Dokshitzer:2005bf,Basso:2006nk}  $C_3=2A_1A_2$, as well as the exact calculations \cite{Zijlstra:1992qd,Moch:1999eb} (see also \cite{Moch:2008fj}) of the two loop coefficient function $c_2(x)$, one finds
eq.(\ref{K-3loop}) does yield an expansion of the form of eq.(\ref{r-expansion-3loop-DIS-ansatz}), which matches as expected the leading ${\cal O}(1/r)$ part (and allows to determine $j_3$).
However,  in the ${\cal O}(r^0)$ part, the exact value of the coefficient of the $a_s^3C_F^2\beta_0\ln^2(1-x)$ term reveals a discrepancy. Indeed, the latter is provided by the 
$C_F^2\, \ln^2(1-x)$ part of  $d_2(x)$ (there is no $C_F^2\frac{\ln^2(1-x)}{1-x}$ part in $d_2(x)$, as correctly predicted by the leading order part of the  ansatz), which is
(see Appendix B) $24C_F^2\, \ln^2(1-x)$, instead of 
 $2A_1^2\, \ln^2(1-x)=32C_F^2\, \ln^2(1-x)$ expected from the ansatz! Equivalently  the ansatz would require (given as input the {\em exact} soft part of $c_1(x)$ (eq.(\ref{c1})), for which no prediction is made) the $C_F^2\, \ln^2(1-x)$ part of $c_2(x)$ to be $64C_F^2\, \ln^2(1-x)$, while the correct result is $60C_F^2\, \ln^2(1-x)$.

 However, the ansatz does make a number of \underline{ correct predictions} (see Appendices A and B), arising essentially from the fact that it correctly implies that $d_2(x)$ contains less power of logarithms for a given color factor then $c_2(x)$. The resulting necessary cancellations\footnote{The fact that the  $\frac{\ln^k(1-x)}{1-x}$ terms occuring in $c_l(x)$ at leading ${\cal O}(1/r)$ order cancel for $l+1\leq k\leq 2l-1$ in the $l$-loop combination $d_l(x)$ (such as $d_2(x)$, $d_3(x)$) which enter $K(x,Q^2)$ at ${\cal O}(a_s^l)$ order was already noticed (in moment space) in \cite{vanNeerven:2001pe}. The present work extend this remark to ${\cal O}(r^0)$ order.} in $d_2(x)$ allow to constrain $c_2(x)$ given $c_1(x)$. The main results are summarized in table (\ref{tab:2loop}).
  
 \begin{table}[hbt]

\begin{center}

\begin{tabular}{|c|c|c|c|c|}
 
 \hline
    & \multicolumn{2}{c|}{$C_F^2$}  &  \multicolumn{2}{c|}{$C_F\beta_0$}  
   
    \\ 
  \hline
  \hline
   $\frac{\ln^3(1-x)}{1-x}$   &   $8$   &   $8$   &   $0$   &   $0$  
   
      \\
   $\bold{\ln^3(1-x)}$   & $\bold{-8}$  & $\bold{-8}$ & $\bold{0}$ & $\bold{0}$ 
   
     \\
   $\frac{\ln^2(1-x)}{1-x}$   &   $-18$   &   $-18$   &   
   $- 2 $   &   $- 2 $  
   
   \\
   $\bold{\ln^2(1-x)}$   &   $\bold{60}$   &   $\bold{64}$   &   
   $\bold{2}$   &   
   $\bold{2}$  
     
   \\
  \hline

\end{tabular}

\end{center}

\caption{\label{tab:2loop} 
Comparison of some exact and predicted 2-loop logarithmic coefficients for the DIS structure
function. Next-to-eikonal results are in boldface. For each color structure, the left column contains the exact results, the right column contains the prediction of the single scale ansatz.}

\end{table}

We note that the ansatz correctly predicts the coefficients of the {\em leading} logarithms for a given color factor\footnote{For leading logarithms, the $C_F(C_A,n_f)$ color factors combine into a {\em single} $C_F\beta_0$ color factor.} in $c_2(x)$, and is also consistent with the general expectation \cite{Kramer:1996iq}  that these coefficients are equal and opposite for the leading  $\frac{\ln^{p}(1-x)}{1-x}$ and $\ln^{p}(1-x)$ terms within each color structure.

  \underline{Another observation}:
 looking at the terms in eq.(\ref{K-3loop}) which contain an explicit $\beta_1$ factor, one finds they are only two at ${\cal O}(a_s^3)$ order: 1) the  $\beta_1c_1(x)$ term and 2) a less obvious contribution contained in $P_2(x)$ (eq.(\ref{P2})). Indeed we have \cite{Dokshitzer:2005bf,Basso:2006nk}:

\begin{equation}
\label{D3}D_3=A_1(B_2^{\delta}-\beta_1)+A_2(B_1^{\delta}-\beta_0)\ .\end{equation} 
Note there is  an additional  $\beta_1$ factor contained in $D_3$. At the ${\cal O}(r^0)$ level, these two  terms proportional to $\beta_1$ thus   contribute  a non-logarithmic piece $-(11C_F+A_1)\beta_1$.
It turns out that this structure is nicely accounted for by assuming that it arises from the total derivative on the right hand side of eq.(\ref{J0-B}). Indeed, setting:

\begin{equation}{\cal B}_0(W^2)=b_1 a_s+a_s^2(-b_1\beta_0\ln(\frac{W^2}{Q^2})+b_2)+a_s^3[b_1\beta_0^2\ln^2(\frac{W^2}{Q^2})-(b_1\beta_1+2\beta_0\, b_2)\ln(\frac{W^2}{Q^2})+b_3]+...\ ,\label{B0-bis}\end{equation}
and

\begin{equation}{\tilde{\cal J}}_{0}\left(W^2\right)={\tilde j}_{02} a_s^2+{\tilde j}_{03} a_s^3+...\label{J0tilde-bis}\ ,\end{equation}
eq.(\ref{J0-B}) yields:

\begin{equation}j_{03}={\tilde j}_{03}-b_1\beta_1-2\beta_0 b_2\ . \label{j03}\end{equation}
Thus the non-logarithmic term in eq.(\ref{r-expansion-3loop-DIS-ansatz}) becomes:

\begin{equation}-j_1\beta_1-2\beta_0\, j_2+j_{03}=-(j_1+b_1)\beta_1-2\beta_0(j_2+b_2)+{\tilde j}_{03}\ ,\end{equation}
where the part proportional to $\beta_1$ indeed reproduces\footnote{It is not possible with the present information, given $j_{03}$, to fix in a unique way ${\tilde j}_{03}$ and $b_2$.} the correct result (since $b_1=11 C_F$). This observation goes beyond what is expected in the large-$\beta_0$ limit.
 We also note   the $j_1\beta_1$ contribution of the ansatz, which arises from the leading $1/r$ term in eq.(\ref{r-expansion-DIS}),  matches the $A_1\beta_1$ term in $D_3$ (paralleling a previous remark concerning the $A_1\beta_0$ term in $D_2$).

\section{Two-scale ansatz}

At the leading order in $1-x$, there is only one scale involved, namely $W^2$.  In next order, however, it is possible that, along with $W^2$, another scale be involved, which would explain the failure of the previous single scale ansatz. We shall {\em assume} this second scale to be given by the `soft' scale ${\widetilde W^2}=(1-x)^2 Q^2$, with the new ansatz:

\begin{align}
\label{r-expansion2-DIS}
\begin{split}
K(x,Q^2)=&\frac{1}{r}\,{\cal J}\left(W^2\right)\,+\,\frac{d\ln \left({\cal F}(Q^2)\right)^2}{d\ln Q^2}\,\delta(1-x)\\
+&\left[\left(\bar{{\cal J}}_{0}\left(W^2\right)-\bar{{\cal S}}_{0}\left({\widetilde W}^2\right)\right)\,\ln(1-x)+{\cal J}_{0}\left(W^2\right)-{\cal S}_{0}\left({\widetilde W}^2\right)\right]\, +{\cal O}\left(r\,\ln^2 r\right)\ ,
\end{split}
\end{align}

Using:
\begin{eqnarray}\bar{{\cal J}}_{0}\left(W^2\right)&=&\bar{j}_{02}a_s^2+a_s^3(-2\bar{j}_{02}\beta_0\ln(\frac{W^2}{Q^2})+\bar{j}_{03})\nonumber\\
&+&a_s^4[3\bar{j}_{02}\beta_0^2\ln^2(\frac{W^2}{Q^2})-(2\beta_1\bar{j}_{02}+3\beta_0\bar{j}_{03})\ln(\frac{W^2}{Q^2})+\bar{j}_{04}]+...
\label{jbar0-3loop}\end{eqnarray}
and

\begin{eqnarray}\bar{{\cal S}}_{0}\left({\widetilde W}^2\right)&=&\bar{s}_{02}a_s^2+a_s^3(-2\bar{s}_{02}\beta_0\ln(\frac{{\widetilde W}^2}{Q^2})+\bar{s}_{03})\nonumber\\
&+&a_s^4[3\bar{s}_{02}\beta_0^2\ln^2(\frac{{\widetilde W}^2}{Q^2})-(2\beta_1\bar{s}_{02}+3\beta_0\bar{s}_{03})\ln(\frac{{\widetilde W}^2}{Q^2})+\bar{s}_{04}]+...
\label{sbar0-3loop}\end{eqnarray}
The  `explicit' $\ln(1-x)$ term in eq.(\ref{r-expansion2-DIS}) thus yields for $x\rightarrow 1$:

\begin{eqnarray}\left(\bar{{\cal J}}_{0}\left(W^2\right)-\bar{{\cal S}}_{0}\left({\widetilde W}^2\right)\right)\,\ln r&=&a_s^2\,(\bar{j}_{02}-\bar{s}_{02})\ln(1-x)\label{explicitlogs-2scales}\\
&+& a_s^3[-2\beta_0(\bar{j}_{02}-2\bar{s}_{02})\ln^2(1-x) +...]\nonumber\\
&+& a_s^4[3\beta_0^2(\bar{j}_{02}-4\bar{s}_{02})\ln^3(1-x) +...]+...\nonumber\end{eqnarray}
where in the last two lines  we have written only the leading logarithms. The new ansatz has more parameters, and one has to go ${\cal O}(a_s^4)$ to get a non-trivial prediction.

\noindent \underline{${\cal O}(a_s^4)$ exact result}:
expanding eq.(\ref{K-loops}) to ${\cal O}(a_s^4)$ one gets:

\begin{eqnarray}
K(x,Q^2)&=&a_s\, P_0(x)+a_s^2[P_1(x)-\beta_0\, c_1(x)]\nonumber\\
&+&a_s^3[P_2(x)-\beta_1\, c_1(x)-\beta_0\, d_2(x)]\label{K-4loop}\\
&+&a_s^4[P_3(x)-\beta_2\, c_1(x)-\beta_1\,  d_2(x)-\beta_0\, d_3(x)]+...\nonumber
\end{eqnarray}
which yields for $x\rightarrow 1$ (see Appendix C), using the expansions of the $c_i(x)$'s provided in \cite{Moch:2008fj} :

\begin{eqnarray}K(x,Q^2)&\sim& \frac{1}{r}\,{\cal J}\left(Q^2(1-x)\right)\nonumber\\
&+&a_s^2(16 C_F^2\ln(1-x)+...)\label{K-4loop-logs}\\
&+&a_s^3(-24 C_F^2\beta_0\ln^2(1-x)+...)\nonumber\\
&+&a_s^4(\frac{88}{3} C_F^2\beta_0^2\ln^3(1-x)+...)+...\nonumber\end{eqnarray}
where in the ${\cal O}(r^0)$ contribution (the last three lines) we have kept only the leading logarithms in each order. Comparing with the corresponding terms (eq.(\ref{explicitlogs-2scales})) in the two-scales ansatz eq.(\ref{r-expansion2-DIS}), one gets the relations:

\begin{eqnarray}\bar{j}_{02}-\bar{s}_{02}&=&16 C_F^2\nonumber\\
2(\bar{j}_{02}-2\bar{s}_{02})&=&24 C_F^2\label{relations}\\
3(\bar{j}_{02}-4\bar{s}_{02})&=&\frac{88}{3} C_F^2\nonumber\end{eqnarray}
Now the first two relations (arising from the ${\cal O}(a_s^2)$ and ${\cal O}(a_s^3)$ contributions) yield:
$\bar{j}_{02}=20 C_F^2$ and $\bar{s}_{02}=4 C_F^2$. However, reporting these values on the left hand side of the third (${\cal O}(a_s^4)$) relation, one gets $3(\bar{j}_{02}-4\bar{s}_{02})=12 C_F^2$, instead of the correct value $\frac{88}{3} C_F^2$! Thus, the two-scale ansatz does not work either. These facts probably indicate failure of threshold resummation at the ${\cal O}(r^0)$ level, in accordance with the analysis of \cite{Laenen:2008gt}.

\noindent\underline{Some correct predictions}:
nevertheless,  the ansatz makes a number of correct predictions (see Appendices A and C), summarized in table (\ref{tab:3loop}). 

\begin{table}[hbt]

\begin{center}

\begin{tabular}{|c|c|c|c|c|c|c|c|c|}
 
 \hline
    & \multicolumn{2}{c|}{$C_F^3$}  &  \multicolumn{2}{c|}{$C_F^2C_A$} &  \multicolumn{2}{c|}{$C_F^2n_f$}   
    & \multicolumn{2}{c|}{$C_F\beta_0^2$}
    \\ 
  \hline
  \hline
   $\frac{\ln^5(1-x)}{1-x}$   &   $8$   &   $8$   &   $0$   &   $0$  &   $0$   &   $0$ 
    &   $0$   &   $0$ 
      \\
   $\bold{\ln^5(1-x)}$   & $ -\bold{8}$  & $-\bold{8}$& $\bold{0}$ & $\bold{0}$ &   $\bold{0}$   &   $\bold{0}$ 
   & $\bold{0}$ & $\bold{0}$
     \\
   $\frac{\ln^4(1-x)}{1-x}$   &   $-30$   &   $-30$   &   
   $-  \frac{220}{9}$   &   $-  \frac{220}{9}$  &   $  \frac{40}{9} $   &   $  \frac{40}{9} $ 
    &   $0$   &   $0$   
   \\
   $\bold{\ln^4(1-x)}$   &   $\bold{92}$   &   $\bold{92}$   &   
   $\bold{\frac{220}{9}}$   &   
   $\bold{\frac{220}{9}}$  &   $\bold{-\frac{40}{9}}$   &   $\bold{-\frac{40}{9}}$ 
    &   $\bold{0}$   &     $\bold{0}$   
   \\
   $\frac{\ln^3(1-x)}{1-x}$   &   $-96\zeta_2-36$   &   $-96\zeta_2-36$   &   $-32\zeta_2+\frac{1732}{9} $   &   $-32\zeta_2+\frac{1732}{9} $  &   $-\frac{280}{9}  $   &   $-\frac{280}{9} $ 
    &   $\frac{4}{3}$   &   $\frac{4}{3}$ 
      \\
   $\bold{\ln^3(1-x)}$   & $ \bold{32\zeta_2-38}$  & $\bold{32\zeta_2-38}$ & $\bold{64\zeta_2-\frac{10976}{27}}$ & $\bold{64\zeta_2-\frac{1156}{3}} $ &   $\bold{\frac{1832}{27} } $   &   $\bold{64 }$ 
   &  $\bold{- \frac{4}{3}}$ & $\bold{- \frac{4}{3}}$
     \\
  \hline

\end{tabular}

\end{center}

\caption{\label{tab:3loop} 
Comparison of some exact and predicted 3-loop logarithmic coefficients for the DIS structure
function. Next-to-eikonal results are in boldface. For each color structure, the left column contains the exact results, the right column contains the prediction of the two-scale ansatz.}

\end{table}

\noindent Similarly to the procedure used at two loop order, one exploits the  cancellations implied by the ansatz in $d_3(x)$ to constrain $c_3(x)$ given $c_2(x)$ and $c_1(x)$.
We observe that the ansatz correctly predicts the coefficients of the {\em leading} logarithms (LL) for a given color factor in $c_3(x)$, and is again  consistent with the general expectation \cite{Kramer:1996iq}  that these coefficients are equal and opposite for the  $\frac{\ln^{p}(1-x)}{1-x}$ and $\ln^{p}(1-x)$ terms within each color structure\footnote{The  $C_A$ and $n_f$ factors combine into a  {\em single} $\beta_0$ factor for leading logarithms, e.g. $C_F^2\beta_0$ for $p=4$ and $C_F\beta_0^2$ for $p=3$.}. We note that these LL predictions depend essentially only upon the validity of the first equation in (\ref{relations}), given the correct leading ${\cal O}(1/r)$ part of the ansatz.

\noindent Moreover, we find that the ansatz also correctly predicts, in term of lower order coefficients,   two {\em subleading} logarithms (the NLL and the NNLL ones)  for the $C_F^3$ color factor
which is associated to the highest  logarithm in  $c_3(x)$: this is a  genuinely new finding of the present approach.
  It is important to note that the prediction of these subleading logarithms in $c_3(x)$ relies on the  knowledge of the {\em exact} soft parts of $c_1(x)$ and $c_2(x)$, or at least of those subleading logarithms in their $x\rightarrow1$ expansion which contribute (see Appendix A) to the relevant subleading logarithms in $c_2(x)\otimes c_1(x)$ and $c_1^{\otimes 3}(x)$.  Since subleading logarithms are involved, these become genuine predictions of the ansatz
only if further parameters are properly adjusted (for instance the second equation in  (\ref{relations}))  so that relevant subleading logarithms in $c_1(x)$ and  $c_2(x)$ are correctly reproduced. However, there is no point of further fixing the ansatz in this way, since not all three relations in (\ref{relations}) can be satisfied anyway, and the ansatz will fail at ${\cal O}(a_s^3)$ as we have seen.

\noindent Finally, the ansatz  at ${\cal O}(1/r)$ predicts there should be no  $C_F ^2(C_A,n_f)\frac{\ln^3(1-x)}{1-x}$ terms in $d_3(x)$. In addition, it predicts that the $C_F ^2(C_A, n_f)\ln^3(1-x)$ terms in $d_3(x)$ should combine into a {\em single} $C_F ^2\beta_0\ln^3(1-x)$ term, which is indeed realized, but with a coefficient $-36/3=-12$  instead of $-88/3$, probably signaling a failure of threshold resummation at the ${\cal O}(r^0)$ level.  The resulting approximate predictions for the $C_F ^2(C_A, n_f)\ln^3(1-x)$ terms in $c_3(x)$ are also displayed in table (\ref{tab:3loop}).

\section{Comment on reference \cite{Laenen:2008ux}}
Setting $Q^2=\mu^2_F$ in the lower limit of the second integral in the large $N$ exponentiation ansatz eq.(37) of \cite{Laenen:2008ux} yields in our notation:

\begin{eqnarray}\ln[\hat{{\cal C}}_2(Q^2,N,\mu^2_F)]&=&(N-independent\  term)+\int_0^1dx\, x^{N-1}\Big[\frac{1}{1-x}B(a_s(rQ^2))\nonumber\\
&+&\int_{\mu^2_F}^{rQ^2}\frac{dk^2}{k^2}P(x,a_s(k^2))+\int_{\tilde{r}Q^2}^{rQ^2}\frac{dk^2}{k^2}\delta P(x,a_s(k^2))\Big]\label{37-0}\ ,\end{eqnarray}
where $ r\equiv\frac{1-x}{x}$,  ${\tilde r}\equiv\frac{(1-x)^2}{x}$ and $P(x,a_s)+\delta P(x,a_s)$ is the time-like (fragmentation) splitting function.
Taking the $\ln Q^2$ derivative of eq.(\ref{37-0}), one gets  for $N\rightarrow\infty$:

\begin{eqnarray}{\hat K}(Q^2,N)\equiv\frac{d\ln[\hat{{\cal C}}_2(Q^2,N,\mu^2_F)]}{d\ln Q^2}&=&\int_0^1dx\, x^{N-1}\Big[P(x,a_s(rQ^2))+\delta P(x,a_s(rQ^2))-\delta P(x,a_s(\tilde{r}Q^2))\nonumber\\
& &+\frac{1}{1-x}\beta(a_s(rQ^2))\frac{dB(a_s(rQ^2))}{da_s}\Big]\label{37}\end{eqnarray}
where we neglected  $N$-independent terms related to the quark form factor (see eq.(\ref{conjecture-0})) outside the integral. Eq.(\ref{37}) 
implies in momentum space  for $x\rightarrow 1$:

\begin{equation}\label{x-ansatz-0}K(x,Q^2)=\Big[P(x,a_s(rQ^2))+\delta P(x,a_s(rQ^2))+\frac{1}{1-x}\beta(a_s(rQ^2))\frac{dB(a_s(rQ^2))}{da_s}\Big]-\delta P(x,a_s(\tilde{r}Q^2))\ .\end{equation}
Setting (cf. eq.(39) in \cite{Laenen:2008ux}):

\begin{equation}\delta P(x,a_s)=\delta C(a_s)\ln(1-x)+\delta D(a_s)\end{equation}
where $\delta C(a_s)=\sum_{i=2}^\infty
\delta C_i a_s^{i}$ and $\delta D(a_s)=\sum_{i=2}^\infty
\delta D_i a_s^{i}$, and using eq.(\ref{P}),
eq.(\ref{x-ansatz-0}) yields:

\begin{eqnarray}K(x,Q^2)&=&\frac{1}{r}{\cal J}(rQ^2)+[ C(a_s(rQ^2))+\delta C(a_s(rQ^2))-\delta C(a_s({\tilde r}Q^2))]\ln(1-x)\label{x-ansatz}\\
&+&D(a_s(rQ^2))+\delta D(a_s(rQ^2))+\beta(a_s(rQ^2))\frac{dB(a_s(rQ^2))}{da_s}-\delta D(a_s({\tilde r}Q^2))\nonumber\ ,\end{eqnarray}
where the $B$-term on the second line arises because in the ansatz of \cite{Laenen:2008ux}  the prefactor of the $B$-term in eq.(\ref{37-0}) is chosen to be  $1/(1-x)$ rather then $1/r=x/(1-x)$ as in the splitting function.
Comparing with eq.(\ref{r-expansion2-DIS}), we deduce that the ansatz of \cite{Laenen:2008ux} is a particular two-scale ansatz, with:

\begin{eqnarray}\bar{{\cal J}}_{0}\left(W^2\right)&=&C\left(a_s\left(W^2\right)\right)+\delta C\left(a_s\left(W^2\right)\right)\label{j0-ansatz}\\
{\cal J}_{0}\left(W^2\right)&=&D\left(a_s\left(W^2\right)\right)+\delta D\left(a_s\left(W^2\right)\right)+\beta(a_s(W^2))\frac{dB(a_s(W^2))}{da_s}\nonumber\end{eqnarray}
and

\begin{eqnarray}\bar{{\cal S}}_{0}\left({\widetilde W}^2\right)&=&\delta C\left(a_s\left({\widetilde W}^2\right)\right)\label{s0-ansatz}\\ 
{\cal S}_{0}\left({\widetilde W}^2\right)&=&\delta D\left(a_s\left({\widetilde W}^2\right)\right)\nonumber\ ,\end{eqnarray}
where we redefined ${\widetilde W}^2={\tilde r}Q^2$.
The approach of \cite{Laenen:2008ux} has the merit to provide a physical justification for the two-scale ansatz. Moreover, this ansatz yields:  $\bar{j}_{02}=-16 C_F^2$ and $\bar{s}_{02}=-32C_F^2$. It is interesting that the first relation in eq.(\ref{relations}) is thus satisfied by the ansatz, but not the other two (we have seen above  that no two-scale ansatz can satisfy all three relations eq.(\ref{relations}) anyway, which presumably signals a failure of threshold resummation at the  ${\cal O}(r^0)$ `next-to-eikonal' level).

\noindent We further observe that the correct predictions of the ansatz of \cite{Laenen:2008ux}, which all concern the {\em leading} ${\cal O}(r^0)$ logarithms for a given color factor, can be obtained for {\em any} two-scale ansatz which satisfy the first relation in eq.(\ref{relations}) (which guarantees the correct leading logarithms in $c_1(x)$),  and such that $\bar{j}_{02}$ and $\bar{s}_{02}$ carry only the $C_F^2$ color factor, as follows from the analysis of section 4. In particular, they are also obtained by the single-scale ansatz ($\bar{s}_{0i}\equiv 0$).

\noindent Another limitation of the ansatz  \cite{Laenen:2008ux}  is that it tries to resum next-to-eikonal logarithms simply by including next-to-eikonal terms in the splitting functions alone. This procedure  is however presumably renormalization scheme dependent beyond leading order, and one should also include\footnote{Actually the ansatz of \cite{Laenen:2008ux} does yield a $B$-term in ${\cal J}_{0}$ (see eq.(\ref{j0-ansatz})). As mentioned in the text, this is due to the use of the prefactor $1/(1-x)$ instead of $1/r$ for the $B$-term in eq.(\ref{37-0}). However, there is no reason why the $B$-term associated to the sub-leading ${\cal J}_{0}$ function should be the same as the one associated to ${\cal J}$ in eq.(\ref{standard-J-coupling}).}  the analogue of the $B$-term in eq.(\ref{37}) at the next-to-eikonal level (see eq.(\ref{analog2})).  In particular, one can check it does not account properly for the large-$\beta_0$ non-logarithmic terms (see eq.(\ref{j02-bis})), which are known \cite{Grunberg:2007nc} to exponentiate, i.e. satisfy the single-scale ansatz.

\section{Concluding remarks}
The present approach shows that threshold resummation either through a single scale, or a two-scale ansatz does not work exactly beyond leading order in $1/r$ (except in the large-$\beta_0$ limit), although these ansaetze do suggest a number of correct predictions for next-to-eikonal logarithms, indicating  only a `partial failure' of threshold resummation at the next-to-eikonal level. Inspired by the analysis of \cite{Laenen:2008gt}, let us assume instead that the physical evolution kernel can be split into the sum of two pieces:

\begin{equation}K(x,Q^2)=K_{exp}(x,Q^2)+K_{nexp}(x,Q^2)\label{K-split}\end{equation}
where only the first piece $K_{exp}(x,Q^2)$ is assumed to have the structure given by (e.g.) the two-scale ansatz eq.(\ref{r-expansion2-DIS}). Then the relations observed in section 3 can be `explained' by assuming that the remainder piece $K_{nexp}(x,Q^2)$  has the following color factors and logarithmic structure in an expansion in $a_s(Q^2)$:

\begin{eqnarray}K_{nexp}(x,Q^2)&=&[k_1 C_F^2\ln(1-x)+k^{'}_1 C_F^2] a_s^2\nonumber\\
&+ &[k_2\, C_F^2\beta_0\ln^2(1-x)+{\cal O}(\ln(1-x))]a_s^3\label{violation}\\
&+ &[k_3\, C_F^2\beta_0^2\ln^3(1-x)+{\cal O}(\ln^2(1-x))]a_s^4+{\cal O}(a_s^5)\nonumber\end{eqnarray}
where the $k_i$'s and $k^{'}_1$ are pure numbers (and eventually $k_1$, and even also $k^{'}_1$, may vanish, i.e. $K_{nexp}(x,Q^2)={\cal O}(a_s^3)$).
Indeed, the cancellation of higher logarithms observed in the combinations $d_i(x)$ is guaranteed by (\ref{violation}), since (as we have seen) these  logarithms cannnot be present in the threshold resummed part $K_{exp}(x,Q^2)$, and are excluded  by assumption from the threshold resummation violating part $K_{nexp}(x,Q^2)$. Further study is required to identify $K_{nexp}(x,Q^2)$, and one may in particular wonder whether the violation of threshold resummation in the physical kernel $K(x,Q^2)$ could be entirely attributed to the coefficient functions $c_i(x)$ ($i\geq 2$) in eq.(\ref{K-4loop}), while the splitting functions $P_i(x)$ themselves would  `exponentiate', i.e. would contribute {\em  only} to $K_{exp}(x,Q^2)$ (which would justify in particular the assumption that $k_1= k^{'}_1=0$, since $c_1(x)$ itself does belong to $K_{exp}(x,Q^2)$, as shown \cite{Grunberg:2007nc} by the large-$\beta_0$ analysis).

Using the results in  \cite{Moch:2008fj}, we have carried  a similar investigation  for the $F_3$ structure function (there is no difference \cite{Moch:2008fj} between the  $F_3$ and $F_1$ coefficient functions up to terms which vanish for $x\rightarrow 1$).  Quite analogous results are obtained (see Appendix D). Moreover the following interesting fact emerged: although the ${\cal O}(r^0)$   next-to-eikonal   logarithms differ between the $F_2$ and $F_3$ coefficient functions (at the difference of the $+$-distributions), we found that, up to three loop, the {\em leading}  next-to-eikonal logarithms are {\em the same} for the $F_2$ and $F_3$  physical evolution kernels (i.e. for the `logarithmic derivative' coefficients $d_i(x)$): see  eq.(\ref{d2-expansion}) and (\ref{d2-F3-expansion}), and
eq.(\ref{d3-expansion-LL}) and (\ref{d3-F3-expansion-LL}). It is natural to wonder whether this feature persists beyond three loop.
 The present approach can also be applied to predict some next-to-eikonal logarithms at four loop order.  A similar study in the Drell-Yan case should also be performed \cite{G}.
\vspace{0.5cm}

{\bf Note added}: after the first version of this paper has been completed, we noticed the paper \cite{Moch:2009mu}, where similar methods are used to deal with the $F_L$ structure function. Moreover, the paper \cite{Moch:2009hr} appeared, which deals with similar issues.


\vspace{0.5cm}

\noindent\textbf{Acknowledgements}

\vspace{0.3cm}

\noindent We thank Andreas Vogt for an early communication prior to publication of the results of \cite{Moch:2008fj} concerning the  $x\rightarrow 1$ expansion of $c_3(x)$.

\appendix

\section{Relevant terms in the expansions of $c_1^{\otimes 2}(x)$,  $c_1^{\otimes 3}(x)$, and $c_2(x)\otimes c_1(x)$  in parametric form}

Writing the soft parts of $c_1(x)$ and $c_2(x)$ as:

\begin{equation}c_1(x)=d_{11}\frac{\ln(1-x)}{1-x}+d_{10}\frac{1}{1-x}+d_{1d}\, \delta(1-x)+b_{11}\ln(1-x)+b_{10}\label{c1-soft}\end{equation}
and:

\begin{eqnarray}c_2(x)&=&d_{23}\frac{\ln^3(1-x)}{1-x}+d_{22}\frac{\ln^2(1-x)}{1-x}+d_{21}\frac{\ln(1-x)}{1-x}+d_{20}\frac{1}{1-x}+d_{2d}\, \delta(1-x)\nonumber\\
& &+b_{23}\ln^3(1-x)+b_{22}\ln^2(1-x)+b_{21}\ln(1-x)+b_{20}\label{c2-soft}\end{eqnarray}
(where the $\ln^p(1-x)/(1-x)$ terms should be understood as $+$-distributions), their convolutions are found to be:

\begin{eqnarray}\label{c1c1-soft}c_1^{\otimes 2}(x)&=&d_{11}^2\frac{\ln^3(1-x)}{1-x}+3 d_{11} d_{10}\frac{\ln^2(1-x)}{1-x}+(-2\zeta_2 d_{11}^2+2d_{11}d_{1d}+2d_{10}^2)\frac{\ln(1-x)}{1-x}+...\nonumber\\
& &+d_{11} b_{11}\ln^3(1-x)+(d_{11} b_{10}+d_{11}^2+2d_{10}b_{11})\ln^2(1-x)\nonumber\\
& &+(-2\zeta_2 d_{11} b_{11}+2d_{11}d_{10}+2d_{10} b_{10}+2d_{1d} b_{11})\ln(1-x)+...\end{eqnarray}

\begin{eqnarray}\label{c1c1c1-soft}c_1^{\otimes 3}(x)&=&\frac{3}{4}d_{11}^3\frac{\ln^5(1-x)}{1-x}+\frac{15}{4} d_{11}^2 d_{10}\frac{\ln^4(1-x)}{1-x}+(-6 \zeta_2 d_{11}^3+6 d_{11} d_{10}^2+3 d_{11}^2 d_{1d})\frac{\ln^3(1-x)}{1-x}+...\nonumber\\
& &+\frac{3}{4}d_{11}^2 b_{11}\ln^5(1-x)+(3 d_{11} d_{10} b_{11}+\frac{3}{4}d_{11}^2 b_{10}+\frac{3}{2}d_{11}^3)\ln^4(1-x)\\
& &+(-6 \zeta_2 d_{11}^2 b_{11}+3 d_{11} d_{10} b_{10}+3 d_{11} d_{1d} b_{11}+6 d_{11}^2 d_{10}+3 d_{10}^2 b_{11})\ln^3(1-x)+...\nonumber
\end{eqnarray}

\begin{eqnarray}\label{c1c2-soft}c_2(x)\otimes c_1(x)&=&\frac{3}{4}d_{23} d_{11}\frac{\ln^5(1-x)}{1-x}+(\frac{5}{4} d_{23} d_{10}+\frac{5}{6} d_{22} d_{11})\frac{\ln^4(1-x)}{1-x}\nonumber\\
& &+(-4 \zeta_2  d_{23}  d_{11}
+ d_{23} d_{1d}+\frac{4}{3} d_{22} d_{10}+d_{21} d_{11})\frac{\ln^3(1-x)}{1-x}+...\nonumber\\
& &+(\frac{1}{4}d_{23} b_{11}+\frac{1}{2}d_{11} b_{23})\ln^5(1-x)\\
& &+(d_{23} d_{11}+\frac{1}{4}d_{23} b_{10}+\frac{1}{3}d_{22} b_{11}+\frac{1}{2}d_{11} b_{22}+ d_{10} b_{23})\ln^4(1-x)\nonumber\\
& &+(-\zeta_2 d_{23} b_{11}-3\zeta_2 d_{11}  b_{23}+ d_{23} d_{10} + d_{22} d_{11}+\frac{1}{3} d_{22} b_{10}+\frac{1}{2} d_{21} b_{11}\nonumber\\
& &+\frac{1}{2} d_{11} b_{21}+d_{10} b_{22}+d_{1d} b_{23})\ln^3(1-x)+...\nonumber
\end{eqnarray}
where for simplicity only  terms relevant for the calculations in Appendices B, C and D have been written down.
In particular these expansions imply, looking at the leading logarithms for a given color structure:

\noindent 1)

\begin{equation}d_2(x)\equiv 2c_2(x)-c_1^{\otimes 2}(x)\supset(2 d_{23}-d_{11}^2)\frac{\ln^3(1-x)}{1-x}+(2 b_{23}-d_{11} b_{11})\ln^3(1-x)\ . \label{d2-soft}\end{equation}
Requiring the coefficients of the logarithms in eq.(\ref{d2-soft}) to vanish then yield the leading logarithm prediction:

\begin{equation}b_{23}=-d_{23}=-\frac{1}{2}d_{11}^2\label{b23}\end{equation}
where the relation
 
\begin{equation}b_{11}=-d_{11}\label{b11}\end{equation}
 has been assumed.

\noindent 2) Writing the soft part of $c_3(x)$ as:

\begin{eqnarray}c_3(x)&=&d_{35}\frac{\ln^5(1-x)}{1-x}+d_{34}\frac{\ln^4(1-x)}{1-x}+d_{33}\frac{\ln^3(1-x)}{1-x}+...\nonumber\\
& &+b_{35}\ln^5(1-x)+b_{34}\ln^4(1-x)+b_{33}\ln^3(1-x)+...\label{c3-soft}\ ,\end{eqnarray}
one gets:

\noindent i)

\begin{equation}d_3(x)\equiv 3\  c_3(x)-3\  c_2(x)\otimes c_1(x)+c_1^{\otimes 3}(x)\supset (3\  d_{35}-\frac{3}{8}d_{11}^3)\frac{\ln^5(1-x)}{1-x}+(3\  b_{35}+\frac{3}{8}d_{11}^3)\ln^5(1-x)\  , \label{d3-soft-1}\end{equation}
where eq.(\ref{b23})  and (\ref{b11}) have been used. Requiring the coefficients of the logarithms in eq.(\ref{d3-soft-1}) to vanish then yield the leading logarithm prediction:

\begin{equation}b_{35}=-d_{35}=-\frac{1}{8}d_{11}^3\label{b35}\ .\end{equation}

\noindent ii) For the quartic logarithms, retaining only those coefficients which contribute to the $C_F^2\beta_0$ color factor, and assuming that $d_{11}$ carries only the $C_F$ color factor,  one gets:

\begin{equation}d_3(x)\supset (3\  d_{34}^{LL}-\frac{5}{2} d_{22}^{LL} d_{11})\frac{\ln^4(1-x)}{1-x}+(3\  b_{34}^{LL}+\frac{5}{2} d_{22}^{LL} d_{11})\ln^4(1-x)\  , \label{d3-soft-2}\end{equation}
where $d_{34}^{LL}$ and $b_{34}^{LL}$ are the parts of $d_{34}$ and $b_{34}$  which belong to the $C_F^2\beta_0$ color factor, and we assumed eq.(\ref{b23}), (\ref{b11}), as well as the relation:

\begin{equation}b_{22}^{LL}=-d_{22}^{LL}\label{b22}\ ,\end{equation}
where $d_{22}^{LL}$ and $b_{22}^{LL}$ are the parts of $d_{22}$ and $b_{22}$  which belong to the $C_F\beta_0$ color factor. Requiring the coefficients of the logarithms in eq.(\ref{d3-soft-2}) to vanish then yield the leading logarithm prediction for the $C_F^2\beta_0$ color factor:

\begin{equation}b_{34}^{LL}=-d_{34}^{LL}=-\frac{5}{6}d_{22}^{LL} d_{11}\label{b34}\ .\end{equation}

\section{Relevant terms in the expansions of $c_2(x)$ and $c_1^{\otimes 2}(x)$ ($F_2$ structure function)}

We use the results, valid for  $x\rightarrow 1$:

\begin{eqnarray}
c_2(x)&=&8\ C_F^2\frac{\ln^3(1-x)}{1-x}-8\ C_F^2\ln^3(1-x)\nonumber\\
& &-18\ C_F^2\frac{\ln^2(1-x)}{1-x}+60\ C_F^2\ln^2(1-x)\nonumber\\
& &-2\ C_F\beta_0\frac{\ln^2(1-x)}{1-x}+2\ C_F\beta_0\ln^2(1-x)\label{c2-expansion}\\
& &+(\frac{16}{3}-8\zeta_2)C_FC_A\frac{\ln(1-x)}{1-x}+(-\frac{34}{3}+24\zeta_2)C_FC_A\ln(1-x)\nonumber\\
& &-(27+32\zeta_2)C_F^2\frac{\ln(1-x)}{1-x}+20C_F^2\ln(1-x)\nonumber\\
& &+\frac{29}{3}C_F\beta_0\frac{\ln(1-x)}{1-x}-\frac{74}{3}C_F\beta_0\ln(1-x)+...\nonumber
\end{eqnarray}
(where we have expressed $n_f$ in term of $\beta_0$ and $C_A$), and

\begin{eqnarray}
c_1^{\otimes 2}(x)&=&16\ C_F^2\frac{\ln^3(1-x)}{1-x}-16\ C_F^2\ln^3(1-x)\nonumber\\
& &-36\ C_F^2\frac{\ln^2(1-x)}{1-x}+96\ C_F^2\ln^2(1-x)\label{c1c1-expansion}\\
& &-(54+64\zeta_2)C_F^2\frac{\ln(1-x)}{1-x}+(-36+64\zeta_2)C_F^2\ln(1-x)+...\nonumber\ .
\end{eqnarray}
These expansions imply  for $r\rightarrow 0$ (using $\frac{1}{1-x}=\frac{1}{r}+1$):

\begin{eqnarray}d_2(x)&=& -4\ C_F\beta_0\frac{\ln^2(1-x)}{r}+24\ C_F^2\ln^2(1-x)\label{d2-expansion}\\
& &+(\frac{32}{3}-16\zeta_2)C_FC_A\frac{\ln(1-x)}{r}+(-12+32\zeta_2)C_FC_A\ln(1-x)\nonumber\\
& &+0\times C_F^2\frac{\ln(1-x)}{r}+(76-64\zeta_2)C_F^2\ln(1-x)\nonumber\\
& &+\frac{58}{3}C_F\beta_0\frac{\ln(1-x)}{r}-30C_F\beta_0\ln(1-x)+...\nonumber\ .
\end{eqnarray}

\section{Relevant terms in the expansions of $c_3(x)$, $c_2(x)\otimes c_1(x)$ and $c_1^{\otimes 3}(x)$ ($F_2$ structure function)}

 For  $x\rightarrow 1$, one gets:

1) \begin{equation}c_3(x)\supset 8\ C_F^3\frac{\ln^5(1-x)}{1-x}-8\ C_F^3\ln^5(1-x)\label{c3-expansion1}\end{equation}

\begin{equation}c_2(x)\otimes c_1(x)\supset 24\ C_F^3\frac{\ln^5(1-x)}{1-x}-24\ C_F^3\ln^5(1-x)\label{c2c1-expansion1}\end{equation}

\begin{equation}c_1^{\otimes 3}(x)\supset 48\ C_F^3\frac{\ln^5(1-x)}{1-x}-48\ C_F^3\ln^5(1-x)\label{c1c1c1-expansion1}\end{equation}
which imply the $C_F^3\frac{\ln^5(1-x)}{1-x}$ and $C_F^3\ln^5(1-x)$ terms cancel in $d_3(x)$ (note this relation concerns {\em leading} logarithms for the color factor $C_F^3$).

2) \begin{equation}c_3(x)\supset -30\ C_F^3\frac{\ln^4(1-x)}{1-x}+92\ C_F^3\ln^4(1-x)\label{c3-expansion2}\end{equation}

\begin{equation}c_2(x)\otimes c_1(x)\supset -90\ C_F^3\frac{\ln^4(1-x)}{1-x}+228\ C_F^3\ln^4(1-x)\label{c2c1-expansion2}\end{equation}

\begin{equation}c_1^{\otimes 3}(x)\supset -180 C_F^3\frac{\ln^4(1-x)}{1-x}+408\ C_F^3\ln^4(1-x)\label{c1c1c1-expansion2}\end{equation}
which imply the $C_F^3\frac{\ln^4(1-x)}{1-x}$ and $C_F^3\ln^4(1-x)$ terms cancel in $d_3(x)$ (note this relation concerns {\em non-leading} logarithms for the color factor $C_F^3$).

3) \begin{equation}c_3(x)\supset - C_F^3(96\ \zeta_2+36)\frac{\ln^3(1-x)}{1-x}+\ C_F^3(32\ \zeta_2-38)\ln^3(1-x)\label{c3-expansion3}\end{equation}

\begin{equation}c_2(x)\otimes c_1(x)\supset -\ C_F^3(288\ \zeta_2+108)\frac{\ln^3(1-x)}{1-x}+C_F^3(224\ \zeta_2-194)\ln^3(1-x)\label{c2c1-expansion3}\end{equation}

\begin{equation}c_1^{\otimes 3}(x)\supset - C_F^3(576\ \zeta_2+216)\frac{\ln^3(1-x)}{1-x}+C_F^3(576\ \zeta_2-468)\ln^3(1-x)\label{c1c1c1-expansion3}\end{equation}
which imply the $ C_F^3\frac{\ln^3(1-x)}{1-x}$ and $ C_F^3\ln^3(1-x)$ terms cancel in $d_3(x)$ (note again this relation concerns {\em non-leading} logarithms for the color factor $C_F^3$).

4) \begin{equation}c_3(x)\supset -\frac{20}{3}\ C_F^2\beta_0\frac{\ln^4(1-x)}{1-x}+\frac{20}{3}\ C_F^2\beta_0\ln^4(1-x)\label{c3-expansion4}\end{equation}

\begin{equation}c_2(x)\otimes c_1(x)\supset -\frac{20}{3}\ C_F^2\beta_0\frac{\ln^4(1-x)}{1-x}+\frac{20}{3}\ C_F^2\beta_0\ln^4(1-x)\label{c2c1-expansion4}\end{equation}
which imply the $C_F^2\beta_0\frac{\ln^4(1-x)}{1-x}$ and $C_F^2\beta_0\ln^4(1-x)$ terms cancel in $ c_3(x)-  c_2(x)\otimes c_1(x)$,  hence also in $d_3(x)$ ($c_1^{\otimes 3}(x)$ has no $C_F^2\beta_0$ color factor). Note this relation concerns {\em leading} logarithms for the color factors $C_F^2(n_f, C_A)$, which combine into a single $C_F^2\beta_0$ color factor in $c_3(x)$ and $c_2(x)\otimes c_1(x)$.

5)\begin{eqnarray}c_3(x)&\supset&  C_F^2[-\frac{280}{9}n_f+(-32\zeta_2+\frac{1732}{9})C_A]\frac{\ln^3(1-x)}{1-x}\nonumber\\
&+& C_F^2[\frac{1832}{27}n_f+(64\zeta_2-\frac{10976}{27})C_A]\ln^3(1-x)\label{c3-expansion5}
\end{eqnarray}

\begin{eqnarray}c_2(x)\otimes c_1(x)&\supset&  C_F^2[-\frac{280}{9}n_f+(-32\zeta_2+\frac{1732}{9})C_A]\frac{\ln^3(1-x)}{1-x}\nonumber\\
&+& C_F^2[\frac{184}{3}n_f+(64\zeta_2-\frac{1112}{3})C_A]\ln^3(1-x)\label{c2c1-expansion5}
\end{eqnarray}
which imply the $C_F^2(n_f, C_A)\frac{\ln^3(1-x)}{1-x}$ terms cancel in $ c_3(x)-  c_2(x)\otimes c_1(x)$ (note this relation concerns {\em non-leading} logarithms for the color factors $C_F^2(n_f, C_A)$),  hence also in $d_3(x)$ ($c_1^{\otimes 3}(x)$ has no $C_F^2(n_f, C_A)$ color factor). Moreover one finds the $C_F^2(n_f, C_A)\ln^3(1-x)$ terms in $d_3(x)$ (but not in $c_3(x)$!) combine into a single term proportional to $C_F^2\beta_0$:
\begin{equation}d_3(x)\supset 0\times C_F^2(n_f, C_A)\frac{\ln^3(1-x)}{1-x}-\frac{88}{3}\ C_F^2\beta_0 \ln^3(1-x)\label{d3-expansion5}\end{equation}

6) \begin{equation}c_3(x)\supset \frac{4}{3}\ C_F\beta_0^2\frac{\ln^3(1-x)}{1-x}-\frac{4}{3}\ C_F\beta_0^2 \ln^3(1-x)\label{c3-expansion6}\end{equation}
which imply , for $r\rightarrow 0$ (using $\frac{1}{1-x}=\frac{1}{r}+1$):

 \begin{equation}c_3(x)\supset \frac{4}{3}\ C_F\beta_0^2\frac{\ln^3(1-x)}{r}+0\times\ C_F\beta_0^2 \ln^3(1-x)\label{c3-expansion6bis}\end{equation}
(note this relation concerns {\em leading} logarithms for the color factors $C_F(n_f^2, n_f C_A, C_A^2)$, which combine into a single $C_F\beta_0^2$ color factor), hence  (since there is no $C_F\beta_0^2$ color factor in $c_2(x)\otimes c_1(x)$ and $c_1^{\otimes 3}(x)$):

 \begin{equation}d_3(x)\supset 4\ C_F\beta_0^2\frac{\ln^3(1-x)}{r}+0\times\ C_F\beta_0^2 \ln^3(1-x)\label{d3-expansion6bis}\end{equation}
 Combining eq.(\ref{d3-expansion5}) and (\ref{d3-expansion6bis}), we thus find the leading logarithms in $d_3(x)$ for $r\rightarrow 0$ are given by:

 \begin{equation}d_3(x)\supset 4\ C_F\beta_0^2\frac{\ln^3(1-x)}{r}-\frac{88}{3}\ C_F^2\beta_0 \ln^3(1-x)\label{d3-expansion-LL}\end{equation}

\section{Relevant terms in the expansions of $c_1(x)$, $c_2(x)$, $c_1^{\otimes 2}(x)$, $c_3(x)$, $c_2(x)\otimes c_1(x)$ and $c_1^{\otimes 3}(x)$ ($F_3$ structure function)}

\underline {Two loop results:} for $x\rightarrow 1$ we have:

 \begin{equation}
\label{c1-F3-usual}
c_1(x)\vert F_3=C_F[\frac{4\ln(1-x)-3}{1-x}-(9+4\zeta_2)\delta(1-x)-4\ln(1-x)+10+...]\ .
\end{equation}
Moreover:

\begin{eqnarray}
c_2(x)\vert F_3&=&8\ C_F^2\frac{\ln^3(1-x)}{1-x}-8\ C_F^2\ln^3(1-x)\nonumber\\
& &-18\ C_F^2\frac{\ln^2(1-x)}{1-x}+52\ C_F^2\ln^2(1-x)\nonumber\\
& &-2\ C_F\beta_0\frac{\ln^2(1-x)}{1-x}+2\ C_F\beta_0\ln^2(1-x)\label{c2-F3-expansion}\\
& &+(\frac{16}{3}-8\zeta_2)C_FC_A\frac{\ln(1-x)}{1-x}+(\frac{14}{3}+8\zeta_2)C_FC_A\ln(1-x)\nonumber\\
& &-(27+32\zeta_2)C_F^2\frac{\ln(1-x)}{1-x}-(16-32\zeta_2)C_F^2\ln(1-x)\nonumber\\
& &+\frac{29}{3}C_F\beta_0\frac{\ln(1-x)}{1-x}-\frac{62}{3}C_F\beta_0\ln(1-x)+...\nonumber
\end{eqnarray}
(where we have expressed $n_f$ in term of $\beta_0$ and $C_A$), and

\begin{eqnarray}
c_1^{\otimes 2}(x)\vert F_3&=&16\ C_F^2\frac{\ln^3(1-x)}{1-x}-16\ C_F^2\ln^3(1-x)\nonumber\\
& &-36\ C_F^2\frac{\ln^2(1-x)}{1-x}+80\ C_F^2\ln^2(1-x)\label{c1c1-F3-expansion}\\
& &-(54+64\zeta_2)C_F^2\frac{\ln(1-x)}{1-x}+(-12+64\zeta_2)C_F^2\ln(1-x)+...\nonumber\ .\end{eqnarray}
These expansions imply  for $r\rightarrow 0$ (using $\frac{1}{1-x}=\frac{1}{r}+1$):

\begin{eqnarray}d_2(x)\vert F_3&=& -4\ C_F\beta_0\frac{\ln^2(1-x)}{r}+24\ C_F^2\ln^2(1-x)\label{d2-F3-expansion}\\
& &+(\frac{32}{3}-16\zeta_2)C_FC_A\frac{\ln(1-x)}{r}+20C_FC_A\ln(1-x)\nonumber\\
& &+0\times C_F^2\frac{\ln(1-x)}{r}-20C_F^2\ln(1-x)\nonumber\\
& &+\frac{58}{3}C_F\beta_0\frac{\ln(1-x)}{r}-22C_F\beta_0\ln(1-x)+...\nonumber\ .
\end{eqnarray}
We observe the leading logarithms on the right hand sides of eq.(\ref{d2-expansion}) and (\ref{d2-F3-expansion}) are {\em identical}.

\noindent\underline {Three loop results:} for $x\rightarrow 1$ we have:

1) \begin{equation}c_3(x)\vert F_3\supset 8\ C_F^3\frac{\ln^5(1-x)}{1-x}-8\ C_F^3\ln^5(1-x)\label{c3-F3-expansion1}\end{equation}

\begin{equation}c_2(x)\vert F_3\otimes c_1(x)\vert F_3\supset 24\ C_F^3\frac{\ln^5(1-x)}{1-x}-24\ C_F^3\ln^5(1-x)\label{c2c1-F3-expansion1}\end{equation}

\begin{equation}c_1^{\otimes 3}(x)\vert F_3\supset 48\ C_F^3\frac{\ln^5(1-x)}{1-x}-48\ C_F^3\ln^5(1-x)\label{c1c1c1-F3-expansion1}\end{equation}
which imply the $C_F^3\frac{\ln^5(1-x)}{1-x}$ and $C_F^3\ln^5(1-x)$ terms cancel in $d_3(x)\vert F_3$ (note this relation concerns {\em leading} logarithms for the color factor $C_F^3$).

2) \begin{equation}c_3(x)\vert F_3\supset -30\ C_F^3\frac{\ln^4(1-x)}{1-x}+84\ C_F^3\ln^4(1-x)\label{c3-F3-expansion2}\end{equation}

\begin{equation}c_2(x)\vert F_3\otimes c_1(x)\vert F_3\supset -90\ C_F^3\frac{\ln^4(1-x)}{1-x}+204\ C_F^3\ln^4(1-x)\label{c2c1-F3-expansion2}\end{equation}

\begin{equation}c_1^{\otimes 3}(x)\vert F_3\supset -180 C_F^3\frac{\ln^4(1-x)}{1-x}+360\ C_F^3\ln^4(1-x)\label{c1c1c1-F3-expansion2}\end{equation}
which imply the $C_F^3\frac{\ln^4(1-x)}{1-x}$ and $C_F^3\ln^4(1-x)$ terms cancel in $d_3(x)\vert F_3$ (note this relation concerns {\em non-leading} logarithms for the color factor $C_F^3$).

3) \begin{equation}c_3(x)\vert F_3\supset - C_F^3(96\ \zeta_2+36)\frac{\ln^3(1-x)}{1-x}+\ C_F^3(96\ \zeta_2-110)\ln^3(1-x)\label{c3-F3-expansion3}\end{equation}

\begin{equation}c_2(x)\vert F_3\otimes c_1(x)\vert F_3\supset -\ C_F^3(288\ \zeta_2+108)\frac{\ln^3(1-x)}{1-x}+C_F^3(288\ \zeta_2-218)\ln^3(1-x)\label{c2c1-F3-expansion3}\end{equation}

\begin{equation}c_1^{\otimes 3}(x)\vert F_3\supset - C_F^3(576\ \zeta_2+216)\frac{\ln^3(1-x)}{1-x}+C_F^3(576\ \zeta_2-324)\ln^3(1-x)\label{c1c1c1-F3-expansion3}\end{equation}
which imply the $ C_F^3\frac{\ln^3(1-x)}{1-x}$ and $ C_F^3\ln^3(1-x)$ terms cancel in $d_3(x)\vert F_3$ (note again this relation concerns {\em non-leading} logarithms for the color factor $C_F^3$).

4) \begin{equation}c_3(x)\vert F_3\supset -\frac{20}{3}\ C_F^2\beta_0\frac{\ln^4(1-x)}{1-x}+\frac{20}{3}\ C_F^2\beta_0\ln^4(1-x)\label{c3-F3-expansion4}\end{equation}

\begin{equation}c_2(x)\vert F_3\otimes c_1(x)\vert F_3\supset -\frac{20}{3}\ C_F^2\beta_0\frac{\ln^4(1-x)}{1-x}+\frac{20}{3}\ C_F^2\beta_0\ln^4(1-x)\label{c2c1-F3-expansion4}\end{equation}
which imply the $C_F^2\beta_0\frac{\ln^4(1-x)}{1-x}$ and $C_F^2\beta_0\ln^4(1-x)$ terms cancel in $ c_3(x)\vert F_3-  c_2(x)\vert F_3\otimes c_1(x)\vert F_3$,  hence also in $d_3(x)\vert F_3$ ($c_1^{\otimes 3}(x)\vert F_3$ has no $C_F^2\beta_0$ color factor). Note this relation concerns {\em leading} logarithms for the color factors $C_F^2(n_f, C_A)$, which combine into a single $C_F^2\beta_0$ color factor in $c_3(x)\vert F_3$ and $c_2(x)\vert F_3\otimes c_1(x)\vert F_3$.

5)\begin{eqnarray}c_3(x)\vert F_3&\supset&  C_F^2[-\frac{280}{9}n_f+(-32\zeta_2+\frac{1732}{9})C_A]\frac{\ln^3(1-x)}{1-x}\nonumber\\
&+& C_F^2[\frac{1640}{27}n_f+(32\zeta_2-\frac{9056}{27})C_A]\ln^3(1-x)\label{c3-F3-expansion5}
\end{eqnarray}

\begin{eqnarray}c_2(x)\vert F_3\otimes c_1(x)\vert F_3&\supset&  C_F^2[-\frac{280}{9}n_f+(-32\zeta_2+\frac{1732}{9})C_A]\frac{\ln^3(1-x)}{1-x}\nonumber\\
&+& C_F^2[\frac{488}{9}n_f+(32\zeta_2-\frac{2696}{9})C_A]\ln^3(1-x)\label{c2c1-F3-expansion5}
\end{eqnarray}
which imply the $C_F^2(n_f, C_A)\frac{\ln^3(1-x)}{1-x}$ terms cancel in $ c_3(x)\vert F_3-  c_2(x)\vert F_3\otimes c_1(x)\vert F_3$ (note this relation concerns {\em non-leading} logarithms for the color factors $C_F^2(n_f, C_A)$),  hence also in $d_3(x)\vert F_3$ ($c_1^{\otimes 3}(x)\vert F_3$ has no $C_F^2(n_f, C_A)$ color factor). Moreover one finds the $C_F^2(n_f, C_A)\ln^3(1-x)$ terms in $d_3(x)\vert F_3$ (but not in $c_3(x)\vert F_3$!) combine into a single term proportional to $C_F^2\beta_0$:
\begin{equation}d_3(x)\vert F_3\supset 0\times C_F^2(n_f, C_A)\frac{\ln^3(1-x)}{1-x}-\frac{88}{3}\ C_F^2\beta_0 \ln^3(1-x)\label{d3-F3-expansion5}\end{equation}

6) \begin{equation}c_3(x)\vert F_3\supset \frac{4}{3}\ C_F\beta_0^2\frac{\ln^3(1-x)}{1-x}-\frac{4}{3}\ C_F\beta_0^2 \ln^3(1-x)\label{c3-F3-expansion6}\end{equation}
which imply , for $r\rightarrow 0$ (using $\frac{1}{1-x}=\frac{1}{r}+1$):

 \begin{equation}c_3(x)\vert F_3\supset \frac{4}{3}\ C_F\beta_0^2\frac{\ln^3(1-x)}{r}+0\times\ C_F\beta_0^2 \ln^3(1-x)\label{c3-F3-expansion6bis}\end{equation}
(note this relation concerns {\em leading} logarithms for the color factors $C_F(n_f^2, n_f C_A, C_A^2)$, which combine into a single $C_F\beta_0^2$ color factor), hence  (since there is no $C_F\beta_0^2$ color factor in $c_2(x)\vert F_3\otimes c_1(x)\vert F_3$ and $c_1^{\otimes 3}(x)\vert F_3$):

 \begin{equation}d_3(x)\vert F_3\supset 4\ C_F\beta_0^2\frac{\ln^3(1-x)}{r}+0\times\ C_F\beta_0^2 \ln^3(1-x)\label{d3-F3-expansion6bis}\ .\end{equation}
 Combining eq.(\ref{d3-F3-expansion5}) and (\ref{d3-F3-expansion6bis}), we thus find the leading logarithms in $d_3(x)\vert F_3$ for $r\rightarrow 0$ are given by:

 \begin{equation}d_3(x)\vert F_3\supset 4\ C_F\beta_0^2\frac{\ln^3(1-x)}{r}-\frac{88}{3}\ C_F^2\beta_0 \ln^3(1-x)\label{d3-F3-expansion-LL}\end{equation}
We observe the right hand sides of eq.(\ref{d3-expansion-LL}) and (\ref{d3-F3-expansion-LL}) are {\em identical}.

\end{document}